\documentclass[aps,preprint,superscriptaddress,nofootinbib]{revtex4}
\usepackage{amsfonts}
\usepackage{amsmath}
\usepackage{bbm}
\usepackage{amssymb}
\usepackage{graphicx}%
\usepackage{footmisc}
\usepackage{bm}
\usepackage{braket}
\usepackage{CJK}
\usepackage{hyperref}
\usepackage{bbm}
\usepackage{cleveref}
\usepackage{longtable,booktabs}

\crefname{equation}{Eq.}{Eqs.}

\newcommand{\Tr}{\text{Tr}}

\begin{document}
\begin{CJK*}{UTF8}{gbsn}
\title{Classical and quantum parts of conditional mutual information for open quantum systems}

\author{Zhiqiang Huang (黄志强)}
\email{hzq@wipm.ac.cn}
\affiliation{State Key Laboratory of Magnetic Resonance and Atomic and Molecular Physics, Innovation Academy for Precision Measurement Science and Technology, Chinese Academy of Sciences, Wuhan 430071, China}
\author{Xiao-Kan Guo (郭肖侃)}
\email{kankuohsiao@whu.edu.cn}
\affiliation{College of Physical Science \& Technology, Bohai University, Jinzhou 121013,  China}

\date{\today}

\begin{abstract}
We study the classical, classical-quantum, and quantum parts of  conditional mutual information in the ``system-environment-ancilla'' setting of open quantum systems. We perform the separation of conditional mutual information by generalizing the classification of correlations of quantum states. The condition for identifying the classical part of conditional mutual information is given by adapting the no-local-broadcasting theorem to this setting, while the condition for classical-quantum part of conditional mutual information is obtained by considering the  multipartite quantum discord and the no-unilocal-broadcasting theorem. For the quantum part of conditional mutual information, we further generalize the characterization of entanglement by quantum discord of state extensions to the multipatite setting, so as to derive a generalized Koashi-Winter-type monogamy equality for conditional mutual information. Our results have explicit dependence on the extensions of environment, which are useful for studying different environmental contributions to the quantum non-Markovianity of open quantum systems.

\end{abstract}


\maketitle

\section{Introduction}\label{INTRO}
The conditional mutual information (CMI) $I(A:B|C)$, which  measures the total information that $A$ and $B$ contain about each other given $C$, is a well-known concept in classical information theory \cite{CT91}.   Although various quantities in quantum information theory can be (re)expressed in terms of CMI \cite{Tucci,CW04}, its applications  in the quantum information tasks begin to prosper only recently. Since the CMI is a multipartite quantity, it is suitable for studying the situations where the multipartite entanglement or correlation becomes important. For example, in the task of quantum state redistribution \cite{DY08,BHOS15}, the CMI quantifies the cost of redistributing of qubits between two quantum systems conditioned on a third quantum system. Other operational interpretations of CMI  include the intrinsic steerability \cite{KWW17},
 the minimum rate of noise for the quantum state deconstruction protocols, the optimal rate of noise for conditional erasure protocols \cite{BBMW18}, and the  optimal rate of secret communication for the conditional one-time pad protocols \cite{SWW20}. In addition to these operational tasks, the CMI also finds applications in the foundational area such as quantum Darwinism \cite{LO19,K21,22a} and in the quantum informational aspects of the AdS/CFT correspondence \cite{KS,DHW16,May21}.

 
 Another important aspect of the quantum CMI is that it is non-negative due to the strong subadditivity of  von Neumann entropy. The particular case of vanishing CMI, $I(A:B|C)=0$, corresponds to the saturation of the data processing inequality for the quantum relative entropies, which is also the quantum Markov chain condition \cite{HJPW04}. Based on these fundamental properties of CMI, several new results have been obtained in recent years: On the one hand, one can deviate a little from the quantum Markov chain condition and obtain the approximate quantum Markov chains \cite{FR15,SW15,BHOS15,SFR16,SR18,KB19,KKB20}. Importantly, the CMI is an upper bound on the fidelity of state reconstruction via the (Petz) recovery maps, thereby bounding the small non-Markovianity present in the approximate Markov chains. A similar bound on the approximate CPTP maps can also be obtained via the CMI \cite{BDW16}.
 On the other hand, by defining the free states as those states satisfying the quantum Markov chain condition, one can develop the resource theory of non-Markovianity for general quantum processes \cite{WSM17,CG19}. The CMI itself can also be exploited to define  a quantifier or a witness of non-Markovianity \cite{HKSK14,HG21,CCMC20}.

In our previous work \cite{HG21}, we have shown particularly that the well-known non-Markovianity quantifier defined via the mutual information $I(A:S)$, known as the Luo-Fu-Song (LFS) measure \cite{LFS}, can have an equivalent form in terms of CMI, $I(A:E|S)$. As the CMI involves three parties, we are able to include the environment or subenvironments into the resulting non-Markovianity quantifier, and consequently we could expect to study the effects of structured environments on the non-Markovianity of general open  quantum processes. The LFS measure, in some sense, quantifies the backflow of correlation from the environment to the system. Since the mutual information $I(A:S)$ contains both the classical  and quantum correlations between $A$ and $S$, it is not clear how to distinguish the classical and quantum contributions to the memory effects leading to the non-Markovian processes (cf. \cite{BMHH20} for a recent discussion on this problem).

In this paper,  we study the problem of distinguishing the classical and quantum  aspects of conditional mutual information with the goal of applying these aspects to the classical and quantum memory effects on the non-Markovian open quantum processes. As is well known, the classical-quantum boundary can be characterized by the no-broadcasting theorem \cite{broadcast,PHH08}. We therefore consider the classical states and the corresponding broadcasting condition for CMI, which identifies the classical part of CMI.
One can also distinguish the classical-quantum and quantum correlations by considering the quantum discord \cite{HV01,OZ02,R3}. We perform a similar analysis for CMI in the setting of open quantum systems, and obtain a similar discord measure defined in terms of CMI, which generalizes the conditional tripartite discord given in \cite{RLB20} by a further minimization.  The vanishing of the minimized conditional tripartite discord defines the classical-quantum CMI, which, when extended to a larger environment $EE'$, is found to be the CMI $I(A:E'|SE)$, meaning that the added   part $E'$ is redundant. This classical-quantum condition for CMI is found to be consistent with the no-unilocal-broadcasting theorem \cite{L10,LS10}, and we find a clear distinction between the classical and the classical-quantum parts of CMI using unilocal broadcasting.
As for the quantum part with nonvanishing minimized conditional tripartite discord, we draw lessons from the method of characterizing of quantum entanglement by the quantum discord after state extensions \cite{L16,XKQK22}, and define a similar entanglement measure using the minimized conditional tripartite discord of  state extensions, which we call the CMI-entanglement. We derive a generalized Koashi-Winter-type monogamy equality for CMI and the CMI-entanglement. When compared to the original Koashi-Winter-type equality \cite{KW04}, we find that the mutual informations and CMI-entanglement involved in this equality contain the extensions of environment.
Therefore, this monogamy equality reveals very different aspects of CMI than the  monogamy inequalities for multipartite quantum discord recently studied in \cite{GHZ21}.  We emphasize that  these results  only hold in our special setting of open quantum systems.

In the next section, we first derive some properties of CMI in  the  setting of  open quantum systems.
 In section \ref{S3}, study the classical, classical-quantum, and quantum parts of CMI in the setting specified in section \ref{S2}.  In section \ref{S4}, we conclude the paper and discuss some further aspects, especially the future application in distinguishing the classical and quantum memory effects in non-Markovian open quantum dynamics.

\section{Conditional mutual information for open quantum systems}\label{S2}
Let $S$ be the quantum system of interest. $S$ interacts with the environment $E$ or any subenvironment  $E$, so that the time evolution of the total $SE$-system is a unitary $U_{SE}$, but the time evolution of $S$ follows open quantum dynamics $\Lambda_S$. We also consider an ancillary system $A$, which is initially maximally entangled with $S$, so as to define the LFS measure of non-Markovianity. Initially, the $AS$-state $\rho^{AS}$ is uncorrelated with the environment state $\rho_0^E$. Then, generally speaking, as the system $S$ evolves  by $\Lambda_{S}$, $S$ would become correlated with $E$. In other words, the environment $E$ would obtain information from the system $S$, or equivalently information would flow into $E$ from $S$. Such an information flow from $S$ to $E$ will reduce or destroy the maximal entanglement between $S$ and $A$, thereby decreasing the mutual information $I(A:S)$. This is simply the monotonicity of the quantum mutual information
under local operations.
But when there is an information backflow from $E$ to $S$, $I(A:S)$ can increase, which happens in the case of non-Markovian open quantum dynamics \cite{LFS}. In this setting, the ancillary system $A$ is used only to keep track of the direction of information flow, but in effect the $A$ can also be considered as a purification of $S$ or as a part of the environment.\footnote{Note that when we interpret $A$ as the purification of $S$, we can simply identify the Hilbert spaces $\mathcal{H}_A=\mathcal{H}_E$.}

Obviously, $I(A:S)$ only contains $S$ and $A$ with the effects from the whole environment $E$ hidden in the open quantum dynamics $\Lambda_S$ on $S$. In our previous work \cite{HG21}, we showed that the LFS non-Markovianity measure can be re-expressed in terms of CMI. This can be easily seen as follows. The quantum mutual information can be decomposed as
\begin{equation}\label{1}
    I(A:SE)=I(A:E|S)+I(S:A),
\end{equation}
where the quantum CMI is
\begin{equation}
    I(A:E|S):=S(\rho_{AS})+S(\rho_{SE})-S(\rho_{S})-S(\rho_{ASE})
\end{equation}
with the von Neumann entropy $S(\rho)=-\text{tr}(\rho\ln \rho)$. Now that we only consider the unitary time evolution of the $SE$-system, there is no information transmission between $A$ and $SE$, and hence we have $\frac{d}{dt}I(A:SE)=0$. With this condition, we can see from \eqref{1} that
\begin{equation}
    \delta I(A:E|S)<0\quad\Leftrightarrow\quad \delta I(A:S)>0.
\end{equation}
Therefore, the non-Markovian condition that $\delta I(A:S)>0$ is equivalent to the condition of decreasing CMI, $\delta I(A:E|S)<0$, along the open quantum dynamics. This way, we are able to build non-Markovianity quantifiers based on the CMI $I(A:E|S)$, now with the explicit $E$-dependence. It is not difficult to see that under the unitary time evolution of the $SE$-system, we have
\begin{equation}
    -\delta I(A:E|S ) \leqslant  I(A:E|S)
\end{equation}
which means that the backflowed CMI cannot be arbitrarily large. This is because if all the correlations between $AS$ and $E$ are backflowed into $S$, one then has $I(AS:E)_{\rho_f}=0$ where $\rho_f$ denotes the final state; since $I(S:E)\geqslant0$,
by \eqref{1} one obtains that $I(A:E|S)_{\rho_f}=0$ and  $    -\delta I(A:E|S ) = I(A:E|S)$, which means that the upper bound of the backflowed CMI under $U_{SE}$ is $I(A:E|S)$.

For later use, let us consider an identity holding only in this particular setup. The initial total $ASE$-state is $\rho^{AS}_0\otimes\rho^{E}_0$, where $\rho_0^{AS}$ is the purified state with $A$ and $S$ being maximally entangled. The subsequent time evolution is given by ${\bf1}_A\otimes U_{SE}$. Since $\rho_0^{AS}$ is a pure state, we have $S(\rho_0^A)=S(\rho_0^S)$. On the other hand, in the initial state $\rho^{AS}_0\otimes\rho^{E}_0$ the environment state $\rho_0^E$ is in tensor product with $\rho^{AS}_0$, so we have $S(\rho_0^{ASE})=S(\rho_0^{AS})+S(\rho_0^E)$ and $S(\rho_0^{SE})=S(\rho_0^S)+S(\rho_0^E)$. Summarizing, we have $S(\rho_{0}^A)=S(\rho_0^{SE})-S(\rho_0^{ASE})$. Because ${\bf1}_A\otimes U_{SE}$ does not change the entropies $S(\rho^A)$, $S(\rho^{SE})$, and $S(\rho^{ASE})$, we have 
\begin{equation}\label{4}
S(\rho^A)=S(\rho^{SE})-S(\rho^{ASE})
\end{equation}
after the time evolution ${\bf1}_A\otimes U_{SE}$ is applied. Using this relation \eqref{4}, we can rewrite \eqref{1} as 
\begin{equation}\label{EQR}
    I(A:E|S)+I(A:S)=2S(A).
\end{equation}
Notice that this relation \eqref{EQR} only holds in the present setting with initial state $\rho^{AS}_0\otimes\rho^{E}_0$ and time evolution ${\bf1}_A\otimes U_{SE}$, and does not hold in general.

We also consider  the following properties of CMI: 
\begin{enumerate}
\item$I(A:E|S)$  {\it does not increase under the local operations and classical communications (LOCC) on } $E$. Indeed, the LOCC on $E$ does not change the $AS$-states, so $\delta I(A:S)=0$. The LOCC on $E$ also will not increase the correlations between $A$ and $SE$. Thus, 
\begin{equation}
    \delta I(A:E|S)=\delta I(A:SE)-\delta I(A:S)\leqslant 0.
\end{equation}
This is expected  from the point of view of resource theory of non-Markovianity \cite{WSM17}.
\item  $I(A:E|S)$ {\it can increase under the LOCC on $S$.} For instance, the CMI for the state $\rho=\sum_i P_i \Pi_i^A\otimes\Pi_i^{E}\otimes\Pi_i^S$, where the $\Pi_i$ are projection operators, is zero, so there is no CMI backflow. But the LOCC on $S$ will change the state to $\rho'=(\sum_i P_i \Pi_i^A\otimes\Pi_i^{E})\otimes\rho_0^S$, for which the CMI is nonvanishing and allows the CMI backflow from $E$ to $S$. 
\item {\it $I(A:E|S)$ does not increase under the extensions of $E$.} Namely, for any extension satisfying $\Tr_{X'}\rho^{ASEX'}=\rho^{ASE}$, we have
\begin{equation}\label{8}
    I(A:EX'|S)_{\rho^{AESX'}}\leqslant   I(A:E|S)_{\rho^{AES}}.
\end{equation}
Indeed, since $\rho^{ASE}(t)={\bf1}\otimes{U}_{SE}(\rho_0^{AS}\otimes \rho_0^E)$, the general form of extension of $E$ can be expressed as $\rho^{ASEX'}(t)=\mathcal{N}_{X\to X'}\circ({\bf1}\otimes{U}_{SE})(\rho_0^{AS}\otimes \rho_0^{EX})$, where $\rho_0^{EX}$ is the extension of $\rho_0^{E}$ and $\mathcal{N}_{X\to X'}$ is any local operation. Because the quantum mutual information does not increase under local operations, we have from \eqref{EQR} that
\begin{equation}\label{P1}
     I(A:EX'S)_{\rho^{ASEX'}(t)}\leqslant   I(A:EXS)_{\rho_0^{AS}\otimes \rho_0^{EX}}=2S(A).
\end{equation}
On the other hand, $\mathcal{N}_{X\to X'}$ also does not influence $ I(A:S)$, so
\begin{equation}\label{P2}
    I(A:S)_{\rho^{ASEX'}(t)} = I(A:S)_{\rho^{AS}(t)} .
\end{equation}
Form \eqref{P1} and \eqref{P2}, we conclude that 
\begin{align}
    I(A:EX'|S)&=  I(A:EX'S)_{\rho^{AESX'}}-  I(A:S)_{\rho^{AESX'}} \leqslant\notag \\
    &\leqslant 2S(A)-  I(A:S)_{\rho^{AS}} =I(A:E|S)
\end{align}
which proves \eqref{8}.
\item {The CMI flow does not change under the extensions of $E$,} i.e., under the unitary time evolution of the $SE$-system, we have
\begin{equation}
    \delta I(A:EX'|S )=  -\delta I(A:S )= \delta I(A:E|S).
\end{equation}
\end{enumerate}
Note that the third property \eqref{8} is proved in our special setting of open quantum systems.

\section{Classical and quantum parts of conditional mutual information}\label{S3}
In this section, we study the classical, classical-quantum, and quantum parts of CMI in the setting of open quantum systems put forward in the last section.

\subsection{Classical part}
The classical part of CMI can be identified by the no-broadcasting theorems. Recall the  no-local-broadcasting theorem for quantum correlations \cite{PHH08}, which  states that  the correlations in a bipartite state $\rho^{AB}$ can be locally broadcast by  $B$ if and only if $\rho^{AB}$ is a classical state $\rho^{AB}=\sum_{ij} p_{i}  \Pi_i^A\otimes \Pi_j^B$.  Then, for the particular classical state $\rho^{ASE}=\sum_{ij} p_{ij} \Pi_{i}^{A} \otimes\Pi_j^{SE}$, the one-side local broadcasting  $\mathcal{E}_{A\to AA'}$ by $A$ broadcasts the correlation to $A'$, that is, 
\begin{equation}
I(A:SE)_\sigma=I(A':SE)_\sigma
\end{equation}
for the reduced states obtained from $\sigma^{AA'ES}=\mathcal{E}_{A\to AA'}(\rho^{ASE})$. Then using \eqref{1} we obtain 
\begin{equation}\label{13}
    I(A:E|S)_\sigma=I(A':E|S)_\sigma.
\end{equation}
This condition tells us that using $A$ and $A'$ one would obtain the same result about the information backflow.
Due to the clear-cut boundary   between the classical and quantum cases as indicated by the no-local-broadcasting theorem, we have the following 

{\it Definition}. For a tripartite state $\rho^{ASE}$ in the ``ancilla-system-environment'' setting with its one-side broadcast state $\sigma^{AA'ES}=\mathcal{E}_{A\to AA'}(\rho^{ASE})$,
the CMI $  I(A:E|S)$ is said to be classical if $I(A:E|S)_\sigma=I(A':E|S)_\sigma$.

Since we have used $A$ as the purification of $S$, we can understand the $A'$ as part of $E$ or as part of  the extension  $EA'\equiv EE'$. We thus say that the correlation in $\rho^{ASE}$ is locally broadcast to  $\rho^{A'SE}$ by the party $A$, and the total state $\sigma^{AA'SE}$ can be reordered to be $\sigma^{ASEA'}\equiv\sigma^{ASEE'}$. In this sense of information backflow, the extension $EA'$ of the environment $E$ does not reveal any new knowledge, meaning that $A'$ is the {\it redundant} extension of $E$.

Let us also recall the no-unilocal-broadcasting theorem \cite{L10,LS10} stating that the correlations in $\rho^{AB}$ can be locally broadcast by party $A$ if and only if $\rho^{AB}$ is a classical-quantum state. In the present setting, we  can apply the above partial theorem to  conclude that the corresponding state allowing unilocal broadcasting of $A$ is a classical-quantum state
\begin{equation}\label{14}
\rho^{ASE}_A=\sum_{i} p_{i} \Pi_{i}^{A} \otimes\rho_i^{SE}.
\end{equation}
This motivates us to consider also the quantum-classical states
\begin{equation}
\rho^{ASE}_E=\sum_{i} p_{i} \rho_{i}^{AS} \otimes\Pi_i^{E}.
\end{equation}
and the unilocal broadcasting of the subenvironment $E$.
However, instead of the classical case as in \eqref{13},  such a unilocal broadcasting of $E$  requires the use of a multipartite generalization of quantum discord, so in the next subsection, we  study the classical-quantum part of CMI directly using the multipartite generalization of quantum discord.
\subsection{Classical-quantum part}
To motivate a multipartite version discord suitable for our present setting, let us recall some reasoning leading to the definition of quantum discord \cite{HV01,OZ02}.
The total correlations between two systems $A$ and $B$ is known to be quantified by the quantum mutual information
   $ I(A:B)=S(\rho^{AB}||\rho^{A}\otimes\rho^{B})$. Given the joint state $\rho^{AB}$, if we perform a POVM measurement on $B$, then we obtain a  classical-quantum state
\begin{equation}
    \rho^{AB}_\text{cq}=\mathcal{M}^B (\rho^{AB})=\sum_i  \Tr_B(M^B_i \rho^{AB})\otimes \Pi_i^B.
\end{equation}
Here, $\mathcal{M}^B$ is a quantum channel, so we can obtain from the monotonicity of mutual information that 
\begin{align}\label{16}
    I(A:B)_{\rho^{AB}}\geqslant   I(A:B)_{\rho^{AB}_\text{qc}}\equiv J(A;B).
\end{align}
By choosing a proper POVM $\{M^B_i\}$, we can obtain the supremum of $J(A;B)$
\begin{equation}\label{CC}
    C(A;B)= \sup_{\{M_i^B\}} J(A;B).
\end{equation}
which defines a measure of classical correlations. The equality in \eqref{16} is not ensured by the sup in \eqref{CC}, and
\begin{equation}
    d(A;B)= I(A:B)-J(A;B)\geqslant0.
\end{equation}
This way,  we can define the quantum discord as 
\begin{equation}
    D(A;B):= I(A:B)-C(A;B).
\end{equation}
When $D(A;B)=0$, we say $\rho^{AB}$ is a classical-quantum state (cf. \eqref{14}).

Now in our setting, we can proceed in a similar way. Firstly, with the help of \eqref{1}, the CMI can be written as 
\begin{equation}
    I(A:E|S)=S(\rho^{ASE}||\rho^A\otimes\rho^{SE} )-S(\rho^{AS}||\rho^A\otimes\rho^{S} ).
\end{equation}
Next, we perform a POVM measurement on $E$ to obtain a classical-quantum state
\begin{equation}
    \rho^{ASE}_\text{cq}=\mathcal{M}^{E} (\rho^{ASE})=\sum_i  \Tr_E(M^{E}_i \rho^{ASE})\otimes \Pi_i^{E}.
\end{equation}
Then by {\it Property 1} in the last section, we have 
\begin{align}\label{DOQCD}
    I(A:E|S)_{\rho^{ASE}} \geqslant I(A;E|S)_{\rho^{ASE}_\text{qc}}\equiv J(A;E|S).
\end{align}
As a CMI, it is obvious that $J(A;E|S)\geqslant 0$. In analogy to $C(A;B)$, we define 
\begin{equation}\label{BLI}
    C(A;E|S):= \sup_{\{M_i^{E}\}} J(A;E|S).
\end{equation}
From \eqref{DOQCD}, we have in general
\begin{equation}\label{24}
    r(A;E|S):= I(A:E|S) -J(A;E|S)\geqslant0.
\end{equation}
Using the chain rule of mutual information and \cref{DOQCD}, one can show the following relations
\begin{align}\label{CROQCD}
    r(A;E|S)&=d(AS;E)-d(S;E),   \notag\\
    r(AA';E|S)&=  r(A;E|SA')+ r(A';E|S).
\end{align}
Finally, for POVM measurements $\{M^{E}_i\}$,
 we define
\begin{equation}\label{RLIBC}
    R(A;E|S):=\min_{\{M^{E}_i\}}  r(A;E|S)= I(A:E|S) -C(A;E|S).
\end{equation}
So we propose the following:

{\it Definition}.
When $R(A;E|S)=0$, we say the corresponding CMI $I(A:E|S) $ is {\it classical-quantum}, otherwise it is {\it quantum}. 

Notice that \eqref{24} is exactly the {\it conditional tripartite discord} defined in  \cite{RLB20}. Here, we have derived it in our special setting of open quantum systems, and further take the minimization to reach \eqref{RLIBC}. Some 3-qubit examples of the  conditional tripartite discord have been given in \cite{RLB20}. To see the behavior of the quantum part of CMI,  let us consider here a new example. Let the  initial state of $AS$ be a maximally entangled state $\ket{\Psi^+}_{AS}=\frac{\sqrt{2}}{2}(\ket{00}+\ket{11})$, and let the initial  state of $E=E_1E_2$ be $\ket{02}_{E_1E_2}$ where $\ket{i}_{E_2},i=0,1,2,$  is a three-level state. Suppose that, after a global unitary evolution, the joint state becomes 
\[\ket{\psi}_{ASE_1E_2}=\sqrt{\frac{u}{2}}(\ket{0000}+\ket{1111})+\sqrt{1-u}\ket{\Psi^+}_{AS}\otimes \ket{02}_{E_1E_2}\]
 with $0\leq u\leq 1$. When $ u=1$, it is easy to see that $I(A:E_1|S)=0$, which means that there cannot be any information backflow from  $E_1$, while from the whole $E$, we obtain $I(A:E_1E_2|S)=\ln 2$. Next, let us consider the measurement basis $\{\cos\theta\ket{00}+e^{i\phi}\sin\theta \ket{11},\sin\theta\ket{00}+e^{-i\phi}\cos\theta\ket{11}\}$ on $E_1E_2$, where $\theta$ and $\phi$ are arbitrary, then after the measurement on $E_1E_2$ we obtain that $C(A;E_1E_2|S)=0$ and $R(A;E_1E_2|S)=\ln 2$. In this case, the $E$ only contributes to the quantum part of CMI, and the information backflow can be completely quantum. When $ u\leqslant 1$, we have the general form of the reduced state
\begin{equation}
    \rho^{ASE_1}=(\frac{u}{2} \Pi_{00}^{AS}+(1-u) \Pi_{\Psi^+}^{AS})\otimes  \Pi_{0}^{E_1} +\frac{u}{2}  \Pi_{11}^{AS}\otimes  \Pi_{1}^{E_1}
\end{equation}
which is a classical-quantum state, thereby $R(A;E_1|S)=0$. If $0< u < 1$, then we have $I(A:E_1|S)=C(A;E_1|S)\neq 0$. For example, if $ u =1/2$, then
\begin{equation}
    I(A:E_1|S)=\frac{1}{4}(\sqrt{5}\ln \frac{2}{3-\sqrt{5}}-3\ln 3+2\ln 2)\approx 0.061.
\end{equation}

Obviously, for classical-quantum states 
\begin{equation}
    \rho_{cq}^{AES}=\sum_i P_i \rho_i^{AS}\otimes \Pi_i^{E}
\end{equation}
we have $R(A;E|S)=0$ thanks to the minimization, but not vice versa.   For any classical-quantum states $ \rho_{cq}^{AES}$,  the corresponding CMI can be broadcast via quantum operation 
\begin{equation}
    \mathcal{N}^{B}_{E\to EE'}(\rho):= \sum_i E_i \rho  E_i^\dagger,
\end{equation}
where  $E_i=\ket{ii}\bra{i}:H^{E}\to H^{E}H^{E'}$. It is easy to see that 
\begin{equation}
    I(A:EE'|S)_{  \mathcal{N}^{B}(\rho_{cq}^{AES})}=I(A:E'|S)_{  \mathcal{N}^{B}(\rho_{cq}^{AES})}=I(A;E|S)_{\rho_{cq}^{AES}},
\end{equation}
since the added $E'$ does not increase the CMI. In this sense, the $E'$ obtained by broadcasting is redundant. We therefore conclude that $C(A;E|S)$ is the part of CMI that can be unilocally broadcast, while $R(A;E|S)$ is the part of CMI that cannot be unilocally broadcast.

We remark that there is an equivalent form of $R(A;E|S)$ based on state extensions. The POVM measurement can be rewritten  in the Kraus representation, $\mathcal{M}^{E}(\cdot)=\sum_{i}K_i(\cdot)K_i^\dag$. The Naimark extension  extends the POVM measurement $\mathcal{M}^{E}$ into a PVM in higher dimensions by an extension $\mathcal{N}_{E\to EE'}$ with the extended state
\begin{align}\label{POVMU}
    \eta^{ASEE'}=\mathcal{N}_{E\to EE'}(\rho^{ASE}).
\end{align}
 Then the state obtained after the POVM measurement is simply the partially traced state $\Tr_{E'}\eta^{ASEE'}=\rho^{ASE}$. It is easy to see that $I(A:E|S)_\rho= I(A:EE'|S)_\eta$. Indeed, from the {\it Property 3} we know that after the extension $E\rightarrow EE'$ the CMI does not increase, i.e., $I(A:E|S)_\rho\geqslant I(A:EE'|S)_\eta$,  while the PVM measurement on $EE'$ considered as a local operation does not increase the CMI by the {\it Property 1}, i.e.,  $ I(A:EE'|S)_\eta\geqslant I(A:E|S)_\rho$, so we have  $I(A:E|S)_\rho= I(A:EE'|S)_\eta$. Thus,
\begin{align}\label{AFRLIBC}
    R(A;E|S)_\rho= I(A:EE'|S)_\eta - C(A;E|S)_\rho.    
\end{align}
In the special case with $C(A;E|S)_\rho=I(A:E|S)_\rho$, i.e., the zero-discord case for classical-quantum CMI, \eqref{AFRLIBC} becomes
\begin{align}\label{29}
    R(A;E|S)_\rho= I(A:EE'|S)_\eta - I(A;E|S)_\rho=I(A:E'|SE)_\eta=0,
\end{align}
where we have used the chain rule for CMI. We therefore see from \eqref{29} that the added $E'$ does not correlate with $A$ from the point of $SE$. In this sense, the added $E'$ can be also understood as redundant. Quite interestingly, this condition \eqref{29} is similar to the strong independence condition in quantum Darwinism \cite{LO19}, if we interpret both  $A$ and $E'$ as subenvironments. 

For later use, we   consider a further extension of $\eta^{ASEE'}$ to $\eta^{ASEE'E''}$, which we choose  to be a purification such that its reduced state $\theta^{AE'S}=\Tr_{EE''}\eta^{AEE'E''S}$ is a classical quantum state. That is to say, we  consider the composite extensions
\begin{align}\label{POVMU1}
    \eta^{ASEE'E''}=\mathcal{N}_{E\to EE'E'' }(\rho^{ASE})=\mathcal{N}^B_{E'\to E'E''}\circ\mathcal{U}_{EE'} (\rho^{AES}\otimes \rho^{E'})
\end{align}
where $\mathcal{U}_{EE'} ((\cdot)\otimes \rho^{E'})=\sum_{ij}K_i (\cdot) K^\dagger_j\otimes \Pi_{ij}^{E'}$. It is easy to verify that the  recovery map
\begin{equation}\label{RMONME}
    \mathcal{R}_{ EE'E''\to E }(\eta):=\Tr_{E'}\circ\mathcal{U}^{-1}_{EE'}  (\sum_i  E_i^\dagger\eta E_i)
\end{equation}
can recovery  exactly the original state $\rho^{ASE}$ from the composite extensions
\begin{equation}
    \mathcal{R}_{ EE'E''\to E }\circ \mathcal{N}_{E\to EE'E'' }(\rho^{ASE})=\rho^{ASE}.
\end{equation}
The state
\begin{equation}
    \theta^{AE'S}=\Tr_{EE''}  \eta^{ASEE'E''}=\sum_{i} \Tr_{E}(K_i \rho^{ASE} K^\dagger_i)\otimes \Pi_{i}^{E'}
\end{equation}
is of the same form as  classical-quantum states $ \rho_{cq}^{AES}$. The property that $ \theta^{AE'S}$ is a classical-quantum state entails $C(A;E'|S)=I(A:E'|S)$, so we have 
\begin{align}\label{AFRLIBC1}
    R(A;E|S)_\rho= I(A:EE'E''|S)_\eta - I(A:E'|S)_\theta     =I(A:EE''|SE')_\eta
\end{align}
where we have used again $I(A:E|S)_\rho=I(A:EE'E''|S)_\eta$ that holds with both extensions and measurements. 

 \subsection{Quantum part}
 Let us further separate the quantum part of CMI. To start with, we continue to consider the state extensions and the characterization of entanglement in these extensions \cite{L16,XKQK22}, namely, an entanglement measure can be defined in terms of quantum discord as
 \begin{equation}\label{EFED}
     \mathcal{E}^a(\rho^{A;E}):=\min_{  \rho^{AEX}} D(A;EX)_{\rho^{AEX}},
   \end{equation}
where $X$ is the state extension of $SE$. In analogy to this, we define in our present setting

{\it Definition}. In the ``ancilla-system-environment'' setting as above, by extending the joint state $\rho^{ASE}$ to $\rho^{ASEX'}$, we define the following quantity
 \begin{equation}\label{LIRTE}
     R_{ex}(A;E|S):=\min_{  \rho^{ASEX'}} R(A;EX'|S)_{\rho^{ASEX'}}
   \end{equation}
where the $ R(A;E|S)$ is given by \eqref{RLIBC}.
   We call this the CMI-entanglement measure.
   
When $X'$ is a trivial extension, we have $ R_{ex}(A;E|S)= R(A;E|S)_\rho$, while for nontrivial extensions we have by {\it Property 3} that
 \begin{equation}
     R(A;E|S)_\rho\geqslant  R_{ex}(A;E|S).
 \end{equation}
 
 To make $R_{ex}(A;E|S)$ a correlation/entanglement measure, so as to derive a generalized Koashi-Winter-type monogamy equality, we show that $R_{ex}(A;E|S)$  satisfies the following desired properties of entanglement monotones:
 
 {\it First}, we show that in our setting of open quantum systems,
  \begin{equation}\label{35}
   R_{ex}(A;E|S)=0 \quad\text{if and only if}\quad \rho^{ASE}=\sum_i P_i \rho^{AS^L}_i\otimes  \rho^{ES^R}_i
  \end{equation}
  where $L,R$ are only labels as conventional in \cite{HJPW04}.
   Indeed, suppose $ R_{ex}(A;E|S)=0$, and we consider the composite state extensions leading to \eqref{AFRLIBC1}, then we have 
\begin{equation}
    I(A:E_XE''_X|SE'_X)_\omega =0,
\end{equation}
where $E_X=EX'$, $\omega^{ASE_XE'_XE''_X}=\mathcal{N}_{E_X\to E_XE'_XE''_X}(\rho^{ASEX'}) $, and $\Tr_{X'} \rho^{ASEX'}=\rho^{ASE}$. Using the similar arguments for separability of states with vanishing squashed entanglement \cite{CW04}, we obtain
\begin{equation}
    \omega^{ASE_XE'_XE''_X}=\sum_i P_i \rho^{AS^L {E'_X}^L}_i\otimes  \rho^{E_XE''_XS^R {E'_X}^R}_i.
\end{equation}
From \cref{POVMU1}, we have $  \eta^{ASE_XE'_X}=\sum_{i}K_i (\rho^{ASEX'}) K^\dagger_i\otimes \Pi_{i}^{E'_X}$. In other words, $A$ and $E'_X$ should not be entangled after the extension, so
\begin{equation}
\omega^{ASE_XE'_XE''_X }=\sum_i P_i \rho^{AS^L}_i\otimes\rho^{E'_X}\otimes  \rho^{E_XE''_XS^R}_i\equiv\sum_i P_i \rho^{AS^L}_i\otimes  \rho^{E_XE''_XS^RE'_X}_i.
\end{equation}
Recovering the original state with \eqref{RMONME} and tracing over the extension $X'$ give back
\begin{align}
    \rho^{AES}=\Tr_{X'} \sum_i P_i \rho^{AS^L}_i\otimes \mathcal{R}_{E_XE'_XE''_X\to E_X}( \rho^{E_XE''_XS^R {E'_X}}_i)    =\sum_i P_i \rho^{AS^L}_i\otimes  \rho^{ES^R}_i.
\end{align}
Conversely, suppose $\rho^{ASE}=\sum_i P_i \rho^{AS^L}_i\otimes  \rho^{ES^R}_i $, then we can choose
\begin{equation}
    \rho^{ASEX}=\sum_i P_i \rho^{AS^L}_i\otimes  \rho^{ES^RC}_i \otimes \Pi_i^D,
\end{equation}
where $X=CD$. We consider also the measurements
\begin{equation}
    \mathcal{M}=\mathcal{M}^{ECD}=\sum_i \mathcal{M}^{EC}_i\otimes \mathcal{M}^D_i
 \end{equation}
 such that
 \begin{equation}
    \mathcal{M}(\rho) ( \rho^{ASEX})= \sum_i P_i \rho^{AS^L}_i\otimes \mathcal{M}^{EC}_i( \rho^{ES^RC}_i) \otimes \Pi_i^D.
 \end{equation}
 Using this type of measurement, we can obtain the corresponding
 \begin{align}
    & r(A:ECD|S)=I(A:ECD|S)_\rho -J(A:ECD|S)_{  \mathcal{M}(\rho)}=\notag \\
     =&I(A:D|S)_\rho-J(A:D|S)_{ \mathcal{M}(\rho)}+\notag \\
     &+I(AS:EC|D)_\rho -J(AS:EC|D)_{  \mathcal{M}(\rho)} -I(S:EC|D)_\rho +J(S:EC|D)_{  \mathcal{M}(\rho)} =0.\label{43}
 \end{align}
 The second line of \eqref{43} vanishes because the measurement $\mathcal{M}$ does not change the $ASD$-part, while the third line vanishes because for any
 $\rho^{ABD}=\sum_i P_i \rho_i^A\otimes \rho_i^B \otimes\Pi_i^D$ its CMI $I(A:B|D)=0$. Finally, by the definition \cref{LIRTE},  we have $ R_{ex}(A;E|S)\leqslant  r(A:ECD|S)=0$, so $ R_{ex}(A;E|S)=0$.

We further observe a special case of extension $X'$ such that  $d(AS;EX')=D(AS;EX')=\mathcal{E}^a(\rho^{AS;E})$, and hence
\begin{align}
    R_{ex}(A;E|S)&\leqslant R(A;EX'|S)\leqslant r(A;EX'|S)=\notag \\
    &=d(AS;EX')-d(S;EX')=D(AS;EX')-d(S;EX')\leqslant \notag \\
    &\leqslant\mathcal{E}^a(\rho^{AS;E})-\mathcal{E}^a(\rho^{S;E})\label{44}
\end{align}
where we have used \cref{CROQCD}. When $A$ is not entangled with $E$ and $AS$ is not globally entangled with $E$, it follows that  $\mathcal{E}^a(\rho^{AS;E})-\mathcal{E}^a(\rho^{A;E})=0$, thereby $  R_{ex}(A;E|S)=0$, which is consistent with \eqref{35}.

We remark that the property \eqref{35} is in some sense a CMI generalization of the separable condition of bipartite states. The particular structure of the states in \eqref{35} is expected to be found for the initial state $\rho_0^{ASE}$. After  time evolution, we learn from the above results that the entanglement structure could change in two ways: one way is to change the $A$-$E$ entanglement, which transfer the $A$-$S$ entanglement to the  $A$-$E$ entanglement; the other way is to change the $A$-$S$-$E$ entanglement, which destroys the  $A$-$S$ entanglement and build global entanglement among $ASE$.

{\it Second}, $R_{ex}(A;E|S)$  is invariant under local unitary operations. This is because  the measurement 
$\{M^{EX'}_i\}$ on the state $\rho^{AESX'}$ is equivalence to the measurement $\{ U_{E} M^{EX'}_i  U_{E}^\dagger\}$ on the transformed state $(U_A\otimes U_S\otimes U_{E})\rho^{AESX'}(U_A\otimes U_S\otimes U_{E})^\dagger$. Considering that the conditional tripartite discord is defined via CMI, this property follows from the local unitary invariance of von Neumann entropy.

{\it Third},  $R_{ex}(A;E|S)$  is invariant  when attaching a pure  ancillary  state $A'$. This is because  the measurement $\{M^{EX'}_i\}$ on the state $\rho^{ASEX'}$ is equivalent to the measurement $\{ M^{EX'}_i  \otimes \Pi_\psi^{E'}\}$ on the attached state $\rho^{ASE}\otimes \Pi_\phi^{A'}\otimes \Pi_\psi^{E'}$.
 
 {\it Fourth}, we show that $R_{ex}(A;E|S)$ is non-increasing under local partial trace in party $A$, i.e.
 \begin{equation}
    R_{ex}(A;E|S)\leqslant  R_{ex}(AA';E|S).
 \end{equation}
 This is because, from \cref{CROQCD}, we have
 \begin{align}
    R_{ex}(AA';E|S)&=\min_{  \rho^{AA'SEX'}}R(AA';EX'|S)_{\rho^{AA'SEX'}} \notag \\
    &=\min_{  \rho^{AA'SEX'}}\min_{\{M_i^{EX'}\}}r(AA';EX'|S)_{\rho^{AA'SEX'}} \notag \\
    & =\min_{  \rho^{AA'SEX'}}\min_{\{M_i^{EX'}\}}\Bigl[  r(A';EX'|SA)_{\rho^{AA'SEX'}}+  r(A;EX'|S)_{\rho^{AA'SEX'}}\Bigr]
      \notag \\
    &\geqslant\min_{  \rho^{AA'SEX'}}\min_{\{M_i^{EX'}\}}  r(A;EX'|S)_{\rho^{AA'SEX'}}\notag \\
      &=\min_{  \rho^{AA'SEX'}}R(A;EX'|S)_{\rho^{AA'SEX'}} =R_{ex}(A;E|S).
   \end{align}  
Similarly, $R_{ex}(A;E|S)$ is non-increasing under local partial trace in party $E$, i.e.
   \begin{equation}
    R_{ex}(A;EE'|S)\geqslant  R_{ex}(A;E|S).
 \end{equation}
 This is because
 \begin{align}
    R_{ex}(A;EE'|S) =\min_{  \rho^{ASEE'X'}}\min_{\{M_i^{EE'X'}\}}r(A;EE'X'|S)_{\rho^{ASEE'X'}} \notag \\
    \geqslant\min_{  \rho^{ASEX''}}\min_{\{M_i^{EX''}\}}  r(A;EX''|S)_{\rho^{ASEX''}}=R_{ex}(A;E|S).
   \end{align}  
 
 {\it Fifth}, $R_{ex}(A;E|S)$ is non-increasing under local operations, i.e.
   \begin{equation}\label{NULO}
    R_{ex}(A;E|S)\geq  R_{ex}(\mathcal{N}^A\otimes \mathcal{N}^{E}(\rho^{ASE})).
 \end{equation}
 Using the Stinespring representation, the local operation  can be realized by adding a pure state ancilla, followed by a global unitary operation and tracing out the ancilla system $\mathcal{N}^S(\cdot)=\Tr_S' \circ \mathcal{U}^{SS'} ((\cdot)\otimes \Pi_0^{S'})$. Then \cref{NULO} is obtained from the preceding three (second, third, and fourth) properties.
 
 {\it Sixth}, $R_{ex}(A;E|S)$ is convex, i.e.
 \begin{equation}
    R_{ex}(\lambda \rho^{ASE}+(1-\lambda )\sigma^{ASE})\leqslant  \lambda  R_{ex}(\rho^{ASE})+(1-\lambda ) R_{ex}(\sigma^{ASE})
 \end{equation}
 for any $0\leqslant\lambda\leqslant 1$.
 Considering  two extensions $ \rho^{ASEX'}$ and $\sigma^{ASEX'}$ of   $ \rho^{ASE}$ and $\sigma^{AES}$ respectively,  we find that the extended state
 \begin{align}
     \tau^{ASEX'Y}\equiv \lambda \rho^{ASEX'}\otimes \Pi_0^{Y}     +(1-\lambda) \sigma^{ASEX'}\otimes \Pi_1^{Y} 
 \end{align}
  gives
 \begin{align}\label{CONVEX}
    &R_{ex}(\lambda \rho^{ASE}+(1-\lambda )\sigma^{ASE})\leqslant r(A;EX'Y|S)_\tau\notag \\
    &=I(A:EX'Y|S)_\tau-J(A:EX'Y|S)_{\mathcal{M}'(\tau)} \notag\\
     &=I(A:EX'|YS)_\tau-J(A:EX'|YS)_{\mathcal{M}'(\tau)} +I(A:Y|S)_\tau-J(A:Y|S)_{\mathcal{M}'(\tau)} \notag\\
      &= \lambda R(A;EX'|S)_\rho +(1-\lambda) R(A;EX'|S)_\sigma.
 \end{align}
 Here, the measurement is of the form 
 \begin{equation}
    \mathcal{M}'=\mathcal{M}^{EX'Y}=\mathcal{M}^{EX'}_0\otimes \mathcal{M}^Y_0+\mathcal{M}^{EX'}_1\otimes \mathcal{M}^Y_1
 \end{equation}
where $\mathcal{M}^{EX'}_0$ is determined by $ R_{ex}(A;E|S)_\rho$,  $\mathcal{M}^{EX'}_1$ is determined by $ R_{ex}(A;E|S)_\sigma$, and $ \mathcal{M}^Y_{i}(\rho)=\Tr_Y(\Pi_i^Y\rho )\otimes \Pi_i^Y $. The reason for the third line of \eqref{CONVEX} is that  the measurement $\mathcal{M}'$ does not change $ASY$, and hence $I(A:Y|S)_\tau=I(A:Y|S)_{\mathcal{M}'(\tau)} $, thereby
\begin{equation}\label{CTT}
    I(A:EX'|YS)_\tau=\lambda I(A:EX'|S)_\rho +(1-\lambda)I(A:EX'|S)_\sigma,
\end{equation}
and similarly for $I(A:EX'|YS)_{\mathcal{M}'(\tau)} $.

With these properties proved, we turn to a generalized Koashi-Winter-type equality for  CMI. For the discord-inspired entanglement measure \eqref{EFED}, the Koashi-Winter equality is proved in \cref{KWE}. 

\begin{figure*}[t]
    \centering
    \includegraphics[width=0.9\textwidth]{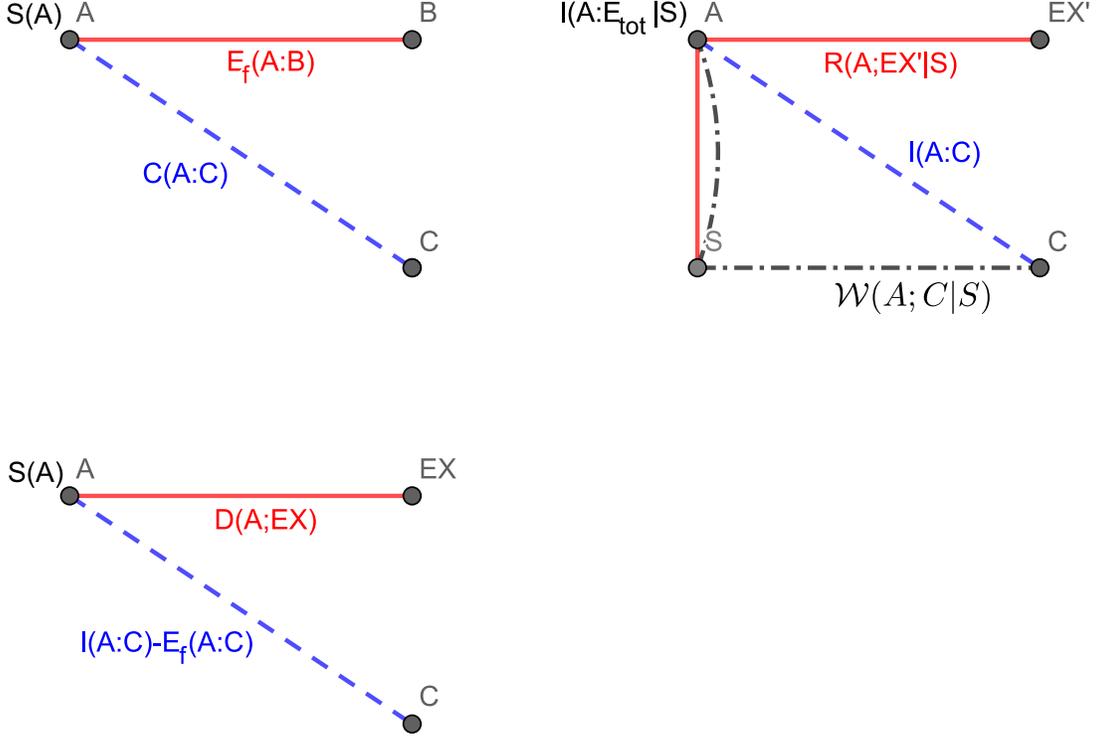}
    \caption{The trade-off relations indicated by the generalized Koashi-Winter-type equality.
    }\label{KWET}
    \end{figure*}

From the definitions \cref{RLIBC,LIRTE} we can write
\begin{equation}\label{REX}
    R_{ex}(A;E|S)=\min_{  \rho^{ASEX'}} R(A;EX'|S)_{\rho^{ASEX'}}=\min_{  \rho^{ASEX'}} \Bigl[I(A:EX'|S)-C(A;EX'|S)\Bigr].
\end{equation}
Let us purify $\rho^{ASEX'}$ to $\phi^{ASEX'C}$.
We observe that 
\begin{align}\label{CMIT}
    I(A:EX'|S)&=I(A:EX'S)-I(A:S)=\notag \\
    &=(S(A)+S(AC)-S(C))-2S(A)+2S(A)-I(A:S)\notag \\
   & =-I(A:C)+I(A:E_\text{tot}|S),
\end{align}
where in the second line we have rewritten the von Neumann entropies based on the pure state $\phi^{ASEX'C}$, and in the third line we have used the special identity \cref{EQR} that holds in the setting with initial state $\rho^{AS}_0\otimes\rho^{E_\text{tot}}_0$ and time evolution ${\bf1}_A\otimes U_{SE_\text{tot}}$. On the other hand, it is easy to show that 
\begin{equation}\label{CCMIT}
    J(A;EX'|S)=S(AS)-S(S)+\sum_i P_i (S(\rho_i^S)-S(\rho_i^{AS}))
\end{equation}
where $P_i=\Tr(M_i^{EX'}\rho^{ASEX'})$, $\rho^{AS}_i=\Tr_{EX'}(M_i^{EX'}\rho^{ASEX'})/P_i$. Since 
\begin{equation}
    \sum_i P_i \rho_i^{AS}= \rho^{AS}=\Tr_C \rho^{ASC}=\Tr_C  \sum_i P'_i \Pi_{\psi_i}^{ASC},
\end{equation}
we can define 
\begin{align}\label{WMIT}
    \mathcal{W}(A;C|S):=S(AS)-S(S)
    +\sup_{\{ P'_i,\ket{\psi_i}\}} \sum_i P'_i (S(\Tr_{AC}\Pi_{\psi_i}^{ASC})-S(\Tr_{C}\Pi_{\psi_i}^{ASC}))
\end{align}
where the  supremum is taken over all decompositions $\{ P'_i,\ket{\psi_i}\}$ with nonnegative probabilities $\rho^{ASC}=\sum_i P'_i \Pi_{\psi_i}^{ASC} $.
As the relative entropy is jointly convex, we have 
\begin{equation}\label{NNOW}
    \mathcal{W}=\sum_i P'_i S(\sigma_i^{AS}||\rho^A\otimes \sigma_i^S)-S(\rho^{AS}||\rho^A\otimes \rho^S)\geqslant 0,
\end{equation}
where $\sigma_i^{AS}=\Tr_{C}\Pi_{\psi_i}^{ASC}$. \cref{NNOW} shows that $ \mathcal{W}(A;C|S)$ is also non-negative. So we have
\begin{equation}\label{EBWAC}
    \mathcal{W}(A;C|S)= C(A;EX'|S).
\end{equation}
Combining \cref{REX,CMIT,EBWAC}, we obtain 
\begin{equation}
    R_{ex}(A;E|S)=I(A:E_\text{tot}|S)-\sup_{ \phi^{ASEX'C}} \Bigl[I(A:C)+\mathcal{W}(A;C|S)\Bigr],
\end{equation}
or 
\begin{equation}\label{GKWEFCMI}
    R(A;EX'|S)_{\rho^{ASEX'}}+I(A:C)+\mathcal{W}(A;C|S)= I(A:E_\text{tot}|S).
\end{equation}
This is the generalized Koashi-Winter equality for CMI. We can see that  the CMI $I(A:E_\text{tot}|S)$ in our setting of open quantum systems can be trade off to the correlations with extended environments. See Fig. \ref{KWET} for a sketch of the trade-off possibilities indicated by the above monogamy equalities.
    
In the present setting, we have an initial  total $ASE_\text{tot}$-state  $\rho^{AS}_0\otimes\rho^{E_\text{tot}}_0$. We  can purify this initial state to $\phi^{ASE_\text{tot}E'_\text{tot}}$. As a special case of \cref{GKWEFCMI}, we can consider $X'$ as a trivial extension $\bar{E}$ such that $E\bar{E}=E_\text{tot}$, and obian
    \begin{equation}
        R(A;E|S)_{\rho^{ASEX'}}+I(A:\bar{E}E'_\text{tot})+\mathcal{W}(A;\bar{E}E'_\text{tot}|S)= I(A:E_\text{tot}|S)
    \end{equation}
    This relation reflects the trade-off between the subenvironment $E$ and the other subenvironment $\bar{E}E'_\text{tot}$.

\section{Concluding Remarks }\label{S4}
We have discussed the classical, classical-quantum, and quantum parts of the CMI, in similar ways of separating the correlation of quantum states. We have put our discussions in the setting of open quantum systems in which we have a CMI  $I(A:E|S)$ relevant to the quantification of non-Markovianity \cite{HG21}. The classical part of CMI is identified by considering broadcasting the $A$-system, and the no-local-broadcasting theorem gives a clear conditional for the classical CMI. The classical-quantum part of  CMI is obtained by generalizing the quantum discord to the CMI, which we found to be a generalization of the conditional tripartite discord to the present multipartite setting. We found that this generalization is consistent with no-unilocal-broadcasting theorem.
As for the quantum part of CMI, we have derived a generalized Koashi-Winter-type monogamy equality for the CMI and also its classical-quantum part as minimized conditional tripartite discord. 

Although these results hold only in our particular setting of open quantum systems, we see the  clear dependence on the extensions of the environment, which would be useful for studying the environmental contributions to classical and quantum memory effects. 
If we use the CMI-based quantifier of non-Markovianity \cite{HKSK14,HG21}, we learn from  the results in this paper that the backflow of CMI is constrained in two ways:  on the one hand, the redundant extensions of the environment do not contribute to the CMI backflow, which happens for both the classical and the classical-quantum parts of CMI; on the other hand, the quantum part of CMI is constrained by 
the monogamy equality, and interestingly this part can be traded off to the extensions of the environment. The effects of these restrictions on quantum non-Markovianity is crucial to understanding almost Markovian phenomenon. 

The distinguishing of the classical and the classical-quantum parts of CMI using unilocal broadcasting also leads us to another way of characterizing the quantum non-Markovianity. As is shown in \cite{BDW16}, the CMI can bound the deviation from complete positivity, which can be related to the resource theory of quantum non-Markovianity based on CP-divisibility. In fact,  in the theory of assignment maps for open quantum systems with initial system-environment correlations, the broadcasting assignment maps also allow the positivity condition to be  discarded \cite{RMA10}. We hope to study these cases in the more general theory of process tensors  where the CMI is also be exploited to define the quantum Markov order \cite{MO}.

We also remark that, in the  discussions of the  classical-quantum parts of CMI,  the  no-unilocal-broadcasting result \cite{L10,LS10} can give a one-to-one correspondence between the vanishing multipartite discord and the classical-quantum states.  To further study the quantum non-Markovianity, we can  change the perspective to the non-Markovianity quantifier based on the quantum Fisher information \cite{SLH15}, as the quantum Fisher information can be broadcast even when the  states  are noncommuting \cite{LSWLO13}. It would be interesting to investigate the relations between these non-Markovianity quantifiers and these no-broadcasting results.
\begin{acknowledgments}
We would like to thank Siddhartha Das for helpful comments.
   ZH  is supported by the National Natural Science Foundation of China under Grant Nos. 12047556, 11725524 and the Hubei Provincial Natural Science Foundation of China under Grant No. 2019CFA003.
\end{acknowledgments}


\appendix
\section{Koashi-Winter-type monogamy relation for  \eqref{EFED}}\label{KWE}
For a pure state $\rho^{ABC}$, the Koashi-Winter monogamy equality is \cite{KW04}
\begin{equation}\label{KWEO}
    E_f(A:B)+C(A;C)=S(A).
\end{equation}
where $E_f(A:B)$ is the entanglement of formation of $\rho^{AB}$ and $C(A;C)$ is the classical correlation.
This equality captures the trade-off between entanglement and classical correlation. It states that the quantum correlation of $A$ to one system $B$ and the classical correlation to the other system $C$ must use up this limited capacity of system $A$ in a mutually exclusive way.

For the discord-inspired entanglement measure \eqref{EFED},  we can derive a similar relation. From the definition \eqref{EFED} we can write
\begin{equation}
    \mathcal{E}^a(\rho^{A;E})=\min_{  \rho^{AEX}} D(A;EX)=\min_{  \rho^{AEX}} \left[I(A:EX)-C(A;EX)\right].
\end{equation}
Let us purify $\rho^{AEX}$ to $\phi^{AEXC}$.  For pure state $\phi^{AEXC}$, we have $S(EX)=S(AC)$ and $S(AEX)=S(C)$. Combining these with \cref{KWEO}, we observe that 
\begin{align}\label{KWEFED}
    \mathcal{E}^a(\rho^{A;E})=\min_{  \phi^{AEXC}} \left[ I(A:EX)-S(A)+E_f(A:C)\right] =S(A)-\sup_{\phi^{AEXC}}\left[I(A:C)-E_f(A:C)\right],
\end{align}
or
\begin{equation}
    D(A;EX)=S(A)-\left[I(A:C)-E_f(A:C)\right].
\end{equation}


\end{CJK*}

\end{document}